\documentclass[conference]{IEEEtran}
\IEEEoverridecommandlockouts
\usepackage{cite}
\usepackage{amsmath,amssymb,amsfonts}
\usepackage{algorithmic}
\usepackage{graphicx}
\usepackage{textcomp}
\usepackage{xcolor}
\usepackage{comment}

\def\BibTeX{{\rm B\kern-.05em{\sc i\kern-.025em b}\kern-.08em
    T\kern-.1667em\lower.7ex\hbox{E}\kern-.125emX}}
\usepackage{bm}

\newcommand{\NN}{\mathbb{N}}

\newcommand{\be}{\begin{equation}}
\newcommand{\ee}{\end{equation}}
\newcommand{\ba}{\begin{array}}
\newcommand{\ea}{\end{array}}

\newcommand{\bb}{{\bm \beta}}

\newcommand{\I}{{\cal I}}
\newcommand{\xiH}{\xi_{{\scriptscriptstyle H}}}
\newcommand{\bxiH}{\bar\xi_{{\scriptscriptstyle H}}}

\newcommand{\Bp}{{\bf p}}

\newcommand{\bx}{{\bf x}}
\newcommand{\bX}{{\bf X}}

\newcommand{\bu}{{\bf u}}
\newcommand{\hD}{\hat\Delta}
\newcommand{\hs}{\hat s}

\newcommand{\hze}{\hat  z_e}

\newcommand{\bup}{{\bf u}^{\scriptscriptstyle \perp}}

\newcommand{\hbup}{\hat {\bf u}^{\scriptscriptstyle \perp}}
\newcommand{\bv}{{\bf v}}

\newcommand{\bj}{{\bf j}}
\newcommand{\Be}{{\bm \epsilon}}
\newcommand{\bE}{{\bf E}}
\newcommand{\bEp}{{\bf E}^{\scriptscriptstyle \perp}}
\newcommand{\bB}{{\bf B}}
\newcommand{\bA}{{\bf A}}
\newcommand{\bAp}{{\bf A}^{\scriptscriptstyle \perp}}

\newcommand{\E}{{\cal E}}
\newcommand{\K}{{\cal K}}

\newcommand{\0}{{\bf 0}}
\newcommand{\bi}{\mathbf{i}}
\newcommand{\bjj}{\mathbf{j}}
\newcommand{\bk}{\mathbf{k}}
\newcommand{\bxp}{{\bf x}^{\scriptscriptstyle \perp}}

\newcommand{\Bep}{{\bm \epsilon}^{\scriptscriptstyle \perp}}
\newcommand{\bjp}{{\bf j}^{\scriptscriptstyle \perp}}
\newcommand{\Bap}{{\bm \alpha}^{\scriptscriptstyle \perp}}
\newcommand{\bXp}{{\bf X}^{\scriptscriptstyle \perp}}

\newcommand{\gammaM}{\gamma^{{\scriptscriptstyle M}}}

\newcommand{\sm}{s_{{\scriptscriptstyle -}}}
\def \sp {s_{{\scriptscriptstyle +}}}
\newcommand{\spm}{s_{{\scriptscriptstyle \pm}}}

\newcommand{\smm}{s_{{\scriptscriptstyle m}}}
\newcommand{\sM}{s_{{\scriptscriptstyle M}}}
\newcommand{\DM}{\Delta_{{\scriptscriptstyle M}}}
\newcommand{\Dm}{\Delta_{{\scriptscriptstyle m}}}
\newcommand{\U}{{\cal U}}

\newcommand{\znd}{z_{{\scriptscriptstyle pd}}}

\begin{document}

\title{A preliminary analysis for efficient laser wakefield acceleration
}

\author{\IEEEauthorblockN{Gaetano Fiore}
\IEEEauthorblockA{\textit{Dip. di Matematica ed Applicazioni, Università di Napoli ``Federico II'',} \\
\textit{and INFN,  Sezione di Napoli,} \\
Napoli, Italy. \\
email: gaetano.fiore@na.infn.it}
}

\maketitle

\begin{abstract}
We propose a preliminary analytical procedure in 4 steps  (based on an improved fully relativistic plane hydrodynamic model)  to tailor the initial density of a cold diluted plasma to the laser pulse profile so as to control wave-breaking (WB) of the plasma wave and maximize the acceleration of small bunches of  electrons self-injected by the first
WB at the density down-ramp. 
\end{abstract}


\section{Introduction}

Laser wake-field acceleration (LWFA)  
\cite{Tajima-Dawson1979,Sprangle1988,EsaSchLee09,TajNakMou17}
is the first and prototypical mechanism of extreme acceleration of charged particles along short distances: injected electrons ``surf" a plasma wave (PW)
driven by a very short laser pulse, e.g. in a diluted supersonic gas jet. 
Dynamics is ruled by Maxwell equations coupled to a  kinetic  theory
for plasma  electrons and ions. Nowadays these equations can be solved via 
more and more 
powerful particle-in-cell (PIC) codes, but since the simulations involve huge costs for each choice of the input data, it is crucial to run them after a preliminary selection of these data based on analyzing simpler models.

Here we use conditions enabling a hydrodynamic description (HD) of the
 impact of a very short and intense laser pulse onto a cold diluted plasma at rest, study
the induced PW and its wave-breaking (WB) 
at  density inhomogeneities  \cite{Daw59}, 
derive preliminary conditions for optimizing  self-injection of small bunches of electrons ($e^-$s) 
in the PW and their LWFA. 
We first adopt an improved fully relativistic plane 
hydrodynamic model \cite{Fio18JPA} (recalled in section \ref{Setup}) whereby the Lorentz-Maxwell and electrons' fluid continuity equations are reduced to the family (parametrized by $Z\!>\!0$) of {\it decoupled pairs}  of Hamilton equations 
 (\ref{heq1})  as long as we can neglect 2-particle collisions, neglect the pulse depletion 
and regard ions as  immobile;  $Z$ pinpoints the infinitesimal layer of $e^-$s having coordinate $z\!=\!Z$ for $\!t\le\! 0$, $\xi=ct\!-\!z$ replaces time $t$ as the independent variable. After the laser-plasma interaction energy is conserved for each $Z$ and  the Jacobian $\hat J$ of the map from Lagrangian to Eulerian coordinates  is 
of the form
\be
\hat J(\xi,Z)=a(\xi,Z)+\xi \, b(\xi,Z),
 \label{lin-pseudoper}
\ee
where $a,b$ are $\xiH$-periodic in $\xi$, and $b$ has zero mean over the period $\xiH(Z)$
given by (\ref{period}). We study the formation of
the PW and its WB where $\hat J\le 0$ (section \ref{hydrobr}), as well as the optimization of the LWFA of self-injected 
$e^-$s (section \ref{WFA}). By causality,  crucial results keep valid  for realistic pulses with a finite spot size (section \ref{Conclu}) and deserve further investigations.

\section{Setup and plane model}
\label{Setup}

\begin{figure*}
\includegraphics[height=3.8cm]{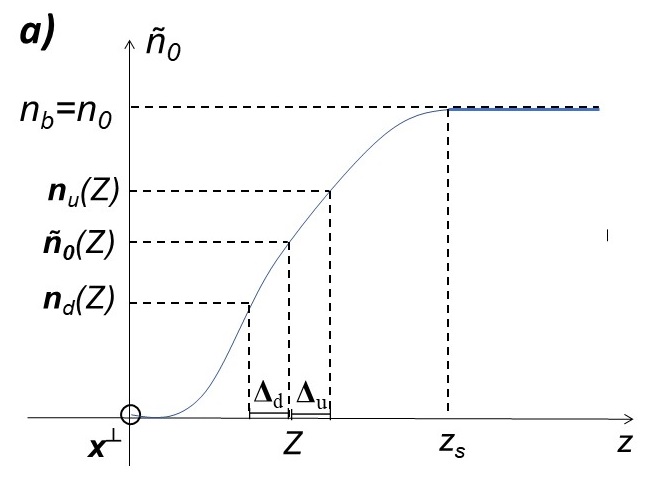}\includegraphics[height=3.8cm]{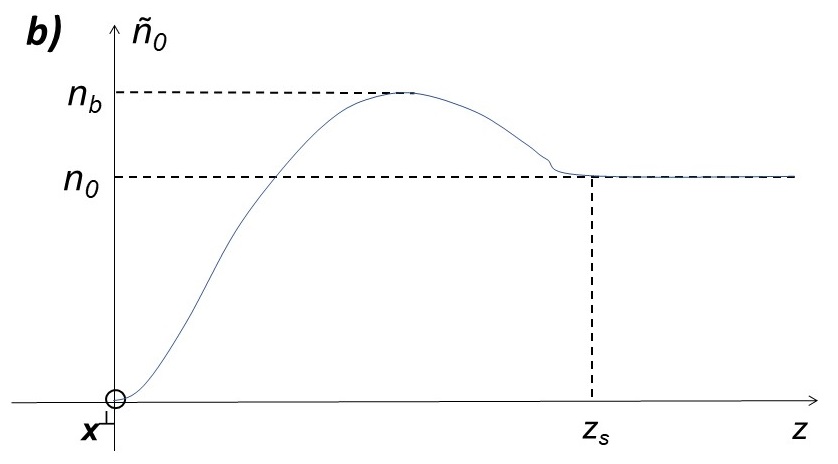}
\includegraphics[height=3.9cm]{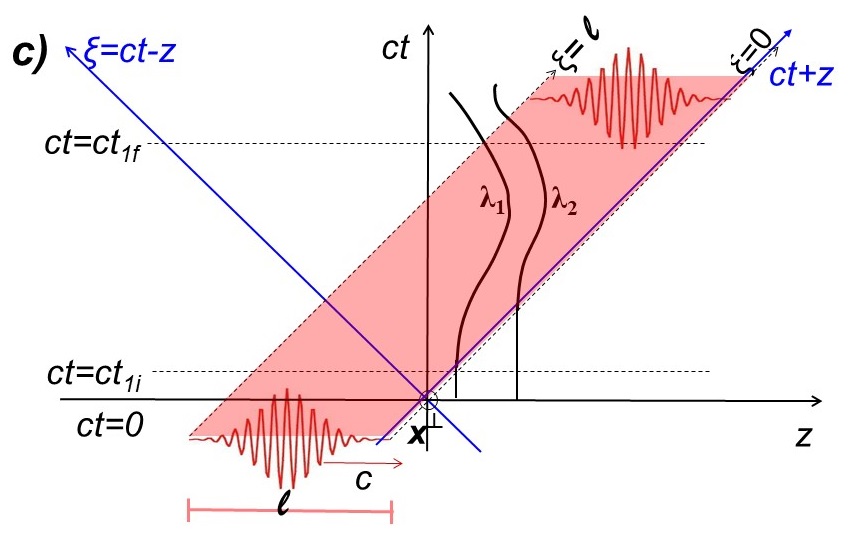}
\caption{a), b): Examples of initial plasma densities of the type
(\ref{n_0bounds}).
c): Projections onto the $z,ct$ plane of sample particle worldlines (WLs) $\lambda_1,\lambda_2$  in Minkowski space \cite{FioCat18}; they intersect the support (pink) of a plane EM wave of total length $l$ moving 
in the positive $z$ direction. 
Since each WL intersects once every hyperplane $\xi\!=$ const (beside every hyperplane $t\!=$ const), we can use $\xi$ rather than $t$ as a parameter along it. While the $t$-instants of intersection with the front and the end of the EM wave (e.g. $t_{1i},t_{1f}$ for $\lambda_1$) depend on the particular WL, the corresponding $\xi$-intstants are the same for all WLs: $\xi_i=0$, $\xi_f=l$.}
\label{fig1}
\end{figure*}

Let the electron fluid  Eulerian density $n_e$, velocity $\bv_e$ fulfill
\be 
\bv_e(0 ,\!\bx)\!=\!\0, \qquad n_e(0,\!\bx)\!=\!\widetilde{n_0}(z), 
 \label{asyc}
\ee
where the initial electron (and proton) density $\widetilde{n_0}(z)$ satisfies
\be 
 \widetilde{n_0}(z) \le  n_b,\qquad \widetilde{n_0}(z)=\left\{\!\!\ba{ll}0 \:\: &\mbox{if }\: z\!\le\! 0, \\
n_0 \:\: &\mbox{if }\: z\!\ge\! z_s \ea\right.
 \label{n_0bounds}
\ee
for some $n_b\!\ge\! n_0\!>\!0$ and $z_s\!>\!0$ (see e.g. Fig. \ref{fig1}): $\widetilde{n_0}(z)$ is bounded by $n_b$ and ends for $z\ge z_s$ with a plateau of height  $n_0$.
Up to section \ref{WFA} we model the  electric and magnetic fields $\bE,\bB$ before the impact ($t\le 0$) 
as a plane wave propagating in the $z$-direction,
\be
\bE (t, \bx)=\bEp (t, \bx)=\Be^{{\scriptscriptstyle\perp}}\!(ct\!-\!z),\quad  \bB=\bB^{{\scriptscriptstyle\perp}}=
\bk\!\times\!\bEp 
      \label{pump}
\ee
($c$ is the speed of light, ${\bf V}^\perp$ denotes the component of vector ${\bf V}$ in the $xy$ plane), where the support of
 $\Be^{{\scriptscriptstyle\perp}}\!(\xi)$ is an interval $0\le\xi\le l$ fulfilling \ 
$l\lesssim  \sqrt{\!\pi mc^2/n_b e^2}$, or more generally  (\ref{Lncond'}), so that
the pulse reaches the plasma at $t\!=\!0$ and overcomes each $e^-$ before the $z$-coordinate of the latter
reaches a negative minimum for the first time ({\it essentially short} pulse).
 $\widetilde{n_0}(z),\Be^{{\scriptscriptstyle\perp}}(\xi)$ make up the {\it input data} of our problem.

\begin{figure}
\includegraphics[width=8.5cm]{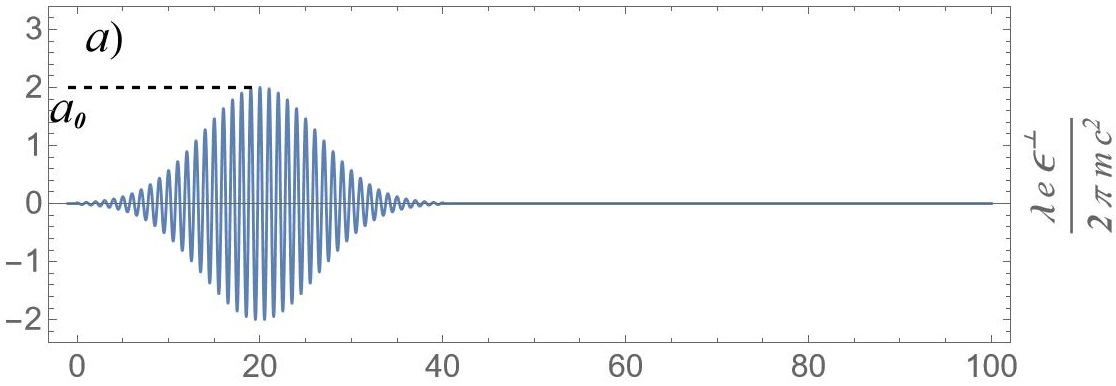}\\
\includegraphics[width=8.6cm]{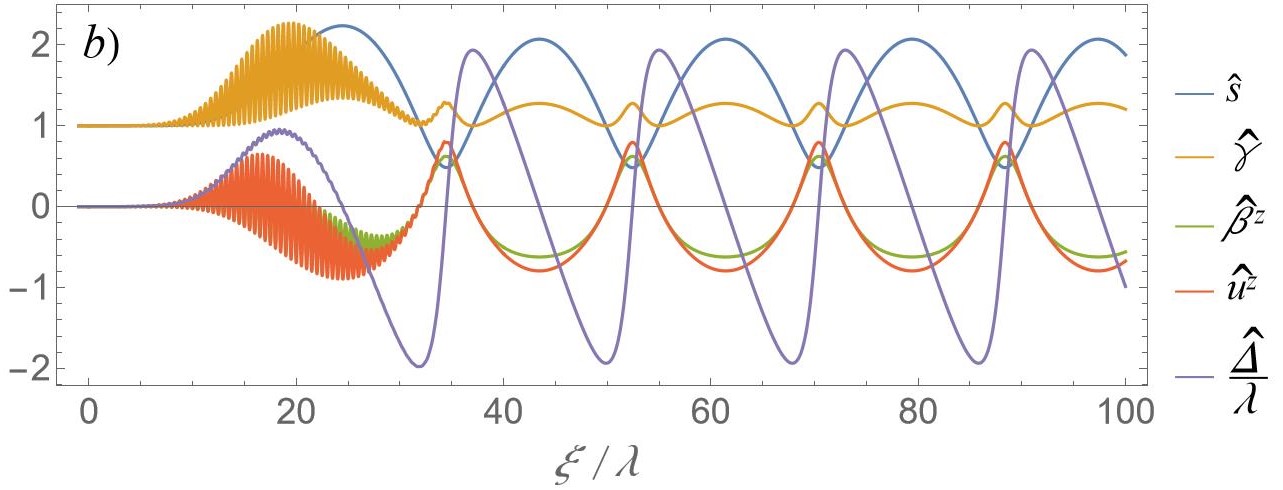}\\
\includegraphics[width=7.7cm]{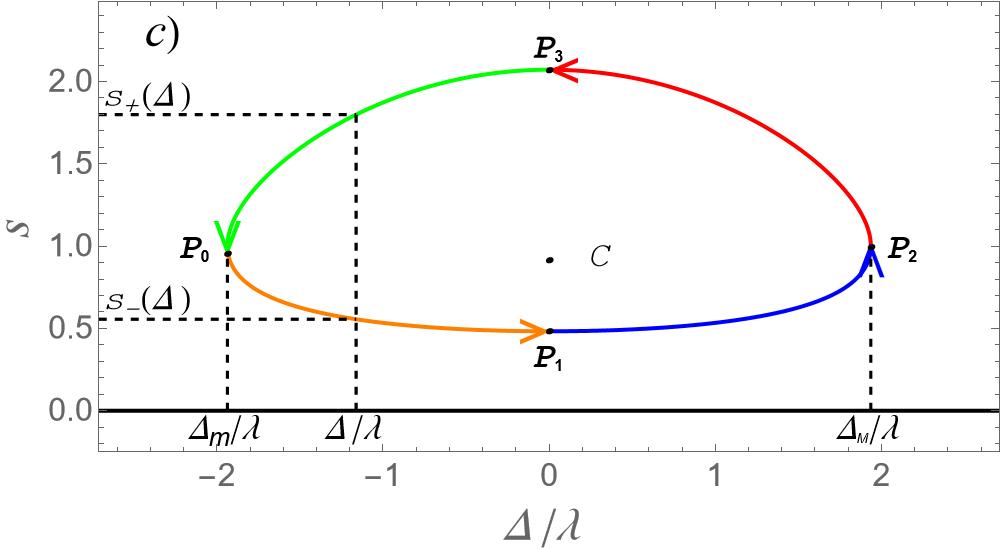}
\caption{
(a)  Normalized gaussian pulse of FWHM $l'
\!=\!10.5\lambda$, linear polarization, 
peak  amplitude $a_0\!\equiv\!\lambda eE^{\scriptscriptstyle \perp}_{\scriptscriptstyle M}/2\pi mc^2\!=\!2$,
as in section III.B of \cite{BraEtAl08}. We consider $l\!=\!40\lambda$ and cut the tails outside $|\xi\!-\!l/2|\!<\!l/2$. 
\ \ (b) \ Corresponding  solution of (\ref{heq1}-\ref{heq2})  if $\widetilde{n_0}(z)\!=\!n_0^j\!\equiv\! n_{cr}/267$ 
($n_{cr}\!=\! \pi mc^2/e^2\lambda^2$ is the critical density); as a result, $E/mc^2\equiv h\!=\!1.28$. Adopting $n_0=n_0^j$ as the density plateau maximizes the LWFA of test electrons, see section \ref{WFA}.{B}. A wavelength $\lambda=0.8\mu$m  leads to a peak  intensity $I\!=\!1.7\!\times\!10^{19}$W/cm$^2$ and
$n_0^j\!=\!6.5\times\! 10^{18}$cm$^{-3}$. 
\ \ (c) \ Corresponding  phase portrait (at $\xi>l$).
}
\label{graphsb}
\end{figure}

The  (Lorentz) equations of motion of an $e^-$ in the unknowns the position \ $\bx(t)$ \ and momentum \ $\Bp(t)=mc\,\bu(t)$ read
\be
\ba{l}
\displaystyle\dot\Bp(t)=-e\bE[t,\bx(t)] - \frac{\Bp(t) \wedge e\bB[t,\bx(t)]  }{\sqrt{m^2c^2\!+\!\Bp^2(t)}} ,\\[6pt]  
\displaystyle
\frac{\dot \bx(t)}c =\frac{\Bp(t) }{\sqrt{m^2c^2\!+\!\Bp^2(t)}}.
\ea
\label{EOM}
\ee
We decompose $\bx\!=\!x\bi\!+\!y\bjj\!+\!z\bk\!=\bxp\!+\!z\bk$, etc, in the Cartesian coordinates of the laboratory frame, and often use the dimensionless variables $\bb\!\equiv\!\bv/c\!=\!\dot\bx/c$,  
\  $\gamma\!\equiv\! 1/\sqrt{1\!-\!\bb^2}\!=\!\sqrt{1\!+\! \bu^2}$, \ 
the 4-velocity $u\!=\!(u^0\!,\bu)\!\equiv\!(\gamma,\gamma \bb)
$, i.e. the dimensionless version of the 4-momentum $p$. As by (\ref{EOM}b) $e^-$ cannot reach the speed of light,  $\tilde \xi(t)\!\equiv\!ct\!-\!z(t)$  grows strictly and admits the inverse $\hat t(\xi)$,
and we can make the change  $t\mapsto \xi\!=\!ct\!-\!z$ of independent parameter  along the worldline (WL)
of $e^-$ (see Fig. \ref{fig1}.c), so that the term $\Bep[ct\!-\!z(t)]$, where the {\it unknown} $z(t)$ is in the argument of the highly nonlinear and rapidly varying $\Bep$, becomes the {\it known} forcing term $\Bep(\xi)$. We denote as $\hat \bx(\xi)$ the position of $e^-$ as a function of $\xi$; it is determined  by $\hat \bx(\xi)=\bx(t)$. More generally we 
denote $\hat f(\xi, \hat \bx)\equiv f\big[\hat t(\xi),  \hat \bx\big]$ (where $c\,\hat t(\xi)\!=\!\xi\!+\!  \hat z(\xi)$) for any given function $f(t,\bx)$,
 abbreviate $\dot f\!\equiv\! df/dt$, $\hat f'\!\equiv\! d\hat f/d\xi$ (total derivatives). 
It is convenient to make also the change of dependent 
variable $u^z\mapsto s$, where the $s$  is the lightlike component of $u$  \cite{Fio18JPA}
\be
s\equiv\gamma\!- u^z=u^-=\gamma(1-\beta^z)=\frac{\gamma}c \frac{d\tilde \xi}{dt}>0;         \label{defs0}
\ee
$\gamma,\bu,\bb$  are the {\it rational} function of $\bu^{{\scriptscriptstyle\perp}}\!\!,s$ 
\be
\gamma\!=\!\frac {1\!+\!\bu^{{\scriptscriptstyle\perp}}{}^2\!\!+\!s^2}{2s}, 
\qquad  u^z\!=\!\frac {1\!+\!\bu^{{\scriptscriptstyle\perp}}{}^2\!\!-\!s^2}{2s}, 
 \qquad  \bb\!=\! \frac{\bu}{\gamma}                                    \label{u_es_e}
\ee
((\ref{u_es_e}) hold also with the carets); $s\!\to\!0$ implies $\gamma,u^z\!\to\!\infty$. 
Replacing $d/dt\mapsto(c s/ \gamma)d/d\xi$ and putting carets  on all variables makes (\ref{EOM}) {\it rational} in the unknowns $\hat\bu^{{\scriptscriptstyle\perp}},\hat s$, in particular  (\ref{EOM}b) becomes $\hat\bx'=\hat\bu/\hat s$. Moreover, $\hat s$ is practically insensitive to fast oscillations
of $\Bep(\xi)$ (as e.g. Fig. \ref{graphsb}.b illustrates).
If $\hat s(\xi)$ vanishes  as $\xi\uparrow \xi_f<\infty$ at least as fast as $\sqrt{\xi_f\!-\!\xi}$, 
then the {\it physical} solution 
vs. $\xi$ is defined only for $\xi<\xi_f$, 
whereas  as a function of $t$ is defined for {\it all} $t<\infty$, because 
$\hat t(\xi_f) =\infty$. 

Passing to the plasma, we denote as $\bx_e(t,\bX)$  the position at time $t$
of the electrons' fluid element $d^3\!X$ initially located at $\bX\!\equiv\!(X,\!Y,\!Z)$, as $\hat \bx_e(\xi,\!\bX)$ the same position    as a function of $\xi$.
 For brevity   we refer to the electrons initially contained: in $d^3\!X$, as the  `$\bX$ electrons';  in a 
region $\Omega$, as the `$\Omega$ electrons'; in the layer between $Z,Z\!+\!dZ$,
as  the `$Z$ electrons'. In the HD
the map $\bx_e(t, \cdot):\bX\mapsto \bx$ must be one-to-one for every $t$;
equivalently,   $\hat\bx_e(\xi, \cdot):\bX\mapsto \bx$ must be one-to-one  for every $\xi$.
The inverses $\bX_e(t,  \cdot):\bx\mapsto \bX$, $\hat \bX_e(\xi,  \cdot):\bx\mapsto \bX$ fulfill 
\be
\bX_e(t,  \bx)=\hat \bX_e(ct\!-\!z,  \bx). \label{clear}
\ee
We can adopt the $x,y$-independent physical observable
\be
\bA\!^{{\scriptscriptstyle\perp}}(t,z) \equiv - c\!\! \int^{t}_{ -\infty }\!\!\!  dt' \: \bEp\!(t',z)
\label{Ap}
\ee
as the transverse component of the EM potential: \ $c\bEp\!\!=\!-\partial_t\bAp$, $\bB\!=\!\bk\!\wedge\!\partial_z\bAp$. Eq. (\ref{pump})  implies for $t\le 0$
\vskip-.5cm
\be
\bAp(t,z)=\Bap(ct\!-\!z),\qquad \Bap(\xi)\!\equiv\! -\!\!\int^{\xi}_{ -\infty }\!\!\!d\eta\:\Bep(\eta).
 \label{Apin}
\ee
Similarly, the displacement  \ $\Delta \bx_e\equiv \bx_e(t,\!\bX\!) -\!\bX$ will 
actually depend only on $t,Z$ [and $\Delta \hat\bx_e\equiv \hat\bx_e(\xi,\bX)\!-\!\bX$ only on $\xi,Z$]
and by causality vanishes  if \  $ct\!\le\! Z$.
The Eulerian electrons' momentum $\Bp_e(t,z)$ obeys equation
(\ref{EOM}), where one has to replace $\bx(t)\mapsto\bx_e(t,\bX)$,
 $\dot \Bp\mapsto d\Bp_e/dt\equiv$ {\it total} derivative; as known, by (\ref{Ap}) 
the transverse part of (\ref{EOM}a)  becomes
$\frac{d\Bp^{{\scriptscriptstyle\perp}}_e}{dt}=\frac ec \frac{d\bAp}{dt}$; 
as $\Bp^{{\scriptscriptstyle\perp}}_e(0,\bx)\!=\!\0\!=\!\Bap(-z)$ 
if $z\!\ge\!0$, this implies
\be
\Bp^{{\scriptscriptstyle\perp}}_e=\frac ec \bAp \qquad\quad 
\mbox{i.e. }\quad \bup_e=\frac e{mc^2} \bAp,         \label{bupExpl}
\ee
which allows to trade $\bup_e$  for $\bAp$ as an unknown. 
The  local conservation \ $n_e\,dz=\widetilde{n_0}\,dZ$ \ of the number of $e^-$
becomes 
\be
 n_e(t,z)=\widetilde{n_0}\!\left[Z_e(t,\!z)\right] \,\partial_z  Z_e(t,z),
\label{n_h}
\ee
and the Maxwell equations \ $\nabla\! \cdot\bE\!-\!4\pi j^0\!=\!\partial_z E^z\!-4\pi e(n_p-n_e)\!=\!0$, \
$ \partial_tE^z/c\!\!+\!4\pi j^z \!\!=\!( \nabla\!\!\wedge\!\bB)^z\!\!=\!0$ ($\bj\!=\!-en_e\bb_e$  is 
the current density) with the initial conditions imply  \cite{Fio14JPA,FioFedDeA14}
\be
E^z(t,\! z)\!=\!4\pi e \!\left\{\!
\widetilde{N}(z)\!-\! \widetilde{N}[Z_e (t,\! z)] \!\right\}\!, \:\:\:
\widetilde{N}(z)\!\equiv\!\!\!\int^z_0\!\!\!\!\!\! d\eta\,\widetilde{n_{0}}(\eta).
 \label{explE}
\ee
Relations (\ref{n_h}-\ref{explE}) allow to express $n_e,E^z$  in terms of 
the assigned $\widetilde{n_0}$
and of the still unknown  $ Z_e(t,z)$ (longitudinal motion); 
thereby  they further reduce the number of unknowns. The remaining ones are $\bA\!^{{\scriptscriptstyle\perp}},\bx_e$ and $u_e^z$, or - equivalently - $s$.

Using the Green function of $\frac 1 {c^2}\partial_t^2\!-\!\partial_z^2$ one reformulates the Maxwell eq. \  $\Box\bAp=4\pi\bjp$ \  and (\ref{Apin}) as the integral eq. \cite{Fio14JPA}
\be
\ba{ll}
\displaystyle\bA\!^{{\scriptscriptstyle\perp}}(t,z)-\Bap(ct\!-\!z)=&\!\!\! \displaystyle\!\!
-  \frac{K}{2} \!\!\!\int \!\!\!   d \eta   d\zeta   \left(\!\frac{n_e\bAp}{\gamma_e}\!\right)\!\left(\eta,\zeta\right)\\[8pt]
&\displaystyle\times \,\theta(\eta)\, \theta\left(ct \!-\!\eta\!-\!|z\!-\!\zeta|\right).  
\ea       \label{inteq1}
\ee
where $K\!\equiv\!\frac{4\pi e^2}{mc^2}$.
The remaining eq.s   (\ref{EOM}) become $\hat\bx^{{\scriptscriptstyle\perp}}_e{}'\!=\!
\hat\bu ^{{\scriptscriptstyle\perp}}_e/\hat s$ and, abbreviating   $v\!\equiv\!\hat\bu^{{\scriptscriptstyle\perp}}_e{}^2\!=\!\big[ \frac{e\hat\bA\!^{{\scriptscriptstyle\perp}}}{mc^2}\big]^2$, 
$\hat\Delta(\xi,Z)\!\equiv\!\hat z_e(\xi,Z)\!-\!Z$, 
\be
\hat\Delta'\!=\displaystyle\frac {1\!+\!v}{2\hat s^2}\!-\!\frac 12, \quad 
\hat s'(\xi,\!Z)\!=\!K\!\left\{\!
\widetilde{N}\!\left[Z\!+\!\hat\Delta\right] \!-\! \widetilde{N}(Z)\! \right\}\!, \label{heq1} 
\ee
with  initial conditions
\ $\hat\Delta(-Z,\!Z)\!=\!0$, $\hat s(-Z,\!Z)\!=\! 1$.  
$\hat s$ cannot vanish anywhere, consistently with 
(\ref{defs0}): if  $\hat s\!\downarrow\! 0$ then  rhs(\ref{heq1}a)  blows up and forces
$\hat\Delta$, and in turn $\hat s$, to abruptly grow again. 
By causality
 $\bA\!^{{\scriptscriptstyle\perp}}(t,z)\!=\!0$ if $ct\!\le\! z$, 
hence $v,\hat\Delta,\hat s-1$ remain zero until $\xi\!=\!0$, 
and we can shift  the initial conditions to
\vskip-.3cm
\be
 \hat \Delta(0,\!Z)\!=\!0,  \qquad\qquad
 \hat s(0,\!Z)\!=\! 1.  \label{heq2}
\ee
\vskip-.1cm
\noindent
Moreover, as  (\ref{inteq1}) is zero for $t\le 0$,  we can still  use (\ref{Apin}),
and by (\ref{bupExpl}) approximate \  $\hat\bu^{{\scriptscriptstyle\perp}}_e\!=\! e\Bap/{mc^2}$, within short time intervals, 
e.g. more precisely in the region $0\!\le\! ct\!-\!z\!\le\! l$, $0\!\le\!\frac{e^2 n_0\lambda}{2 mc^2}(ct+z)\!\ll\! 1$
\cite{FioDeNAkhFedJov23}; 
$\hat\bu ^{{\scriptscriptstyle\perp}}_e$ and the forcing term $v$ thus
become {\it known} functions of $\xi$ (only), and 
 (\ref{heq1})  a family parametrized by $Z$ of {\it decoupled ODEs}.
For every $Z$ they have the form of Hamilton equations \ $q'=\partial \hat H/\partial p$, $p'=-\partial \hat H/\partial q$  of a 1-dim system: \  $\xi,\hat\Delta, -\hat s$  play the role of $t,q,p$, and the Hamiltonian is {\it rational} in $\hat s$ and reads \cite{FioDeN16} 
\begin{IEEEeqnarray}{c}
\hat H( \hat \Delta, \hat s,\xi;Z)\equiv \gamma(\hat s;\xi)+ \U( \hat \Delta;Z), \qquad\qquad
    \label{hamiltonian} \\[6pt]
\gamma(s;\!\xi) \!\equiv\frac{s^2\!+\!1\!+\!v(\xi)}{2s},\quad
\frac{\U( \Delta;\!z)}K \!\equiv\! \int^{z \!+\!  \Delta}_z\!\!\!\!\!\!\!\!\!\!\!d\zeta\,\widetilde{N}(\zeta)
- \widetilde{N}\!(z)  \Delta;
\nonumber                            
\end{IEEEeqnarray}
$\gamma\!-\!1$, $\U$ act as kinetic, potential energy in $mc^2$ units.
We  can easily solve (\ref{heq1}-\ref{heq2}) in the unknown $\hat P\equiv(\hat\Delta,\hat s)$ numerically, or by quadrature for 
$\xi\!\ge\! l$. Finally, \ $\hat\bx^{{\scriptscriptstyle\perp}}_e{}'\!=\!
\hat\bu ^{{\scriptscriptstyle\perp}}_e/\hat s$  is solved by
\vskip-.2cm
\be
\hat\bx^{\scriptscriptstyle \perp}_e(\xi,\bX)-\bXp=\!\int^\xi_0\!\!\! d\eta \,\frac{\hat\bu^{\scriptscriptstyle \perp}_e(\eta)}{\hat s(\eta,Z)}.         \label{hatsol'}
\ee
The PW emerges as a collective effect  passing to the Eulerian description, 
see e.g. Fig.  \ref{Worldlinescrossings-new}.
We shall say that a pulse is
\be
\mbox{{\it essentially short (ES) w.r.t. $\widetilde{n_0}$ } \ if}\quad 
\hs(\xi,Z)\ge 1, 
 \label{Lncond'}
\ee
for all $\xi\!\in\![0,\!l]$, $Z\!\ge\! 0$. \ 
Sufficient conditions on the input data for ES 
pulses are given in \cite{FioDeAFedGueJov22}.
ES pulses simplify the control of the PW and its WB, allowing to compute
apriori bounds on  $\hD ,\hs,\hat J$, e.g. $\Delta_u,\Delta_d$ such that
$\Delta_u(z)\!\ge\!\hD (\xi,z) \!\ge\!\Delta_d(z)$.

\noindent
As in Fig. \ref{graphsb}a, below we assume the pulse is a ES slowly modulated monochromatic wave (SMMW); 
this makes $v(l)\ll 1$. 
Approximating $v(l)\!=\!0$, it follows $v(\xi)\!=\!0$,  $\bup(\xi)\!=\!\0$ and by (\ref{hatsol'}) \
$\hat\bx^{\scriptscriptstyle \perp}_e(\xi,\!\bX)\!=$const (purely longitudinal motion)  for $\xi\!\ge\! l$.

\subsection{Electrons' motion after the laser-plasma interaction ($\xi>l$)}
\label{after}

For $\xi\!>\!l$ energy $H$ is conserved along the $Z$-solution;  if $Z\!>\!0$ the path in $P\equiv (\Delta,s)$ phase space is a cycle around $
(0,1)$. Its points $P$ solve
$\hat H(\Delta,s;Z)\!=\!h(Z)\!\equiv\!1\!+\!\!\int^l_0\!\!d\xi v'(\xi)/\hat s(\xi,Z)$; 
\be
P_0\!\equiv\!(\Dm,\!1), \quad
 P_1\!\equiv\!(0,\!\smm), \quad
P_2\!\equiv\!(\DM,\!1), \quad
 P_3\!\equiv\!(0,\!\sM) \nonumber
\ee
minimize/maximize $\Delta$ or $s$, see Fig. \ref{graphsb}.c. The periodic motion can be determined by quadrature;  the period is
\be
\xi_{{\scriptscriptstyle H}}(Z)=2\int^{\DM(Z)}_{\Dm(Z)}\!\!\!\!\!\!\!\!\!d\Delta\, \frac{\bar\gamma(\Delta;Z)}{\sqrt{ \bar\gamma^2(\Delta;Z)-1}}, \quad  \bar\gamma\equiv h\!-\!\U.                        \label{period}
\ee
$\DM\!>\!0,\Dm\!<\!0$ solve the equation
$\U(\Delta;Z)\!=\!h(Z)\!-\!1$. The points $P$ with given  $\Delta\!\in\![\Dm,\!\DM]$ are
$P=(\Delta,\spm)$, where $\spm\!\equiv\! \bar\gamma\pm\sqrt{\bar\gamma^2\!-\!1}$,
and  $\sM(Z)=\sp(0,\!Z)$, $\sm(Z)=\sm(0,\!Z)$.

\subsection{Special case: \ $\widetilde{n_0}(Z)\equiv n_0=$ const}
\label{specialcase}

This implies $Z$-independence of: $\U\!=\!M\Delta^2/2$ ($M\!\equiv\! Kn_0$), (\ref{heq1}-\ref{heq2}), its solution 
$P\equiv \big(\Delta,s\big)$ plotted in Fig. \ref{graphsb} ($s'=M\Delta$),  $\DM\!=\!-\Dm\!=\!\sqrt{2(h\!-\!1)/M}$,
$\hat J\!=\!1$. Each $Z$ layer of $e^-$  is a copy of the same 
relativistic harmonic oscillator.  (\ref{period}) becomes
\be
\bar \xi_{{\scriptscriptstyle H}}\!\big(n_0,\!h\big)
=8\sqrt{\!\frac{h\!+\!1}{2M}}\left[\E(\alpha)-\frac
{\K(\alpha)}{h\!+\!1}\right], 
\quad \alpha\equiv\sqrt{\!\frac{h\!-\!1}{h\!+\!1}} ; \label{period0}
\ee
$\K,\E$ are the complete elliptic integrals of the 1st, 2nd kind.
($\bar \xi_{{\scriptscriptstyle H}}/c$ resp. reduces to 
$t_{{\scriptscriptstyle H}}^{{\scriptscriptstyle nr}}\!\equiv\!\sqrt{\!\frac{\pi m}{n_0 e^2}}$, $t_{{\scriptscriptstyle H}}^{{\scriptscriptstyle ur}}\!\simeq\! \frac{15}{8}\sqrt{\!\frac{h\pi m}{2n_0 e^2}}$ 
 in the nonrelativistic  limit $h\!\to\!1$ and the  ultrarelativistic one $h\!\to\!\infty$.)
 $\hat Z_e(\xi,\!z;\!n_0)\!=\!z\!-\!\Delta(\xi;\!n_0)$, and the Eulerian fields are travelling waves with speed $c$, e.g. \ $n_e(t,\!z)=\frac{n_0}2 
\left[1\!+\! \frac{1\!+\!v(ct\!-\!z)}{s^2(ct\!-\!z)}\right]$.
It turns out that the solution $\hat P(\xi,Z)$ of (\ref{heq1}-\ref{heq2}) for generic (not too small) $\widetilde{n_0}$
is very close to $P(\xi; n_0)$ with  $n_0\!=\!\widetilde{n_0}(Z)$.

\section{Hydrodynamic regime up to wave-breaking}
\label{hydrobr}

\begin{figure}
\includegraphics[width=9cm]{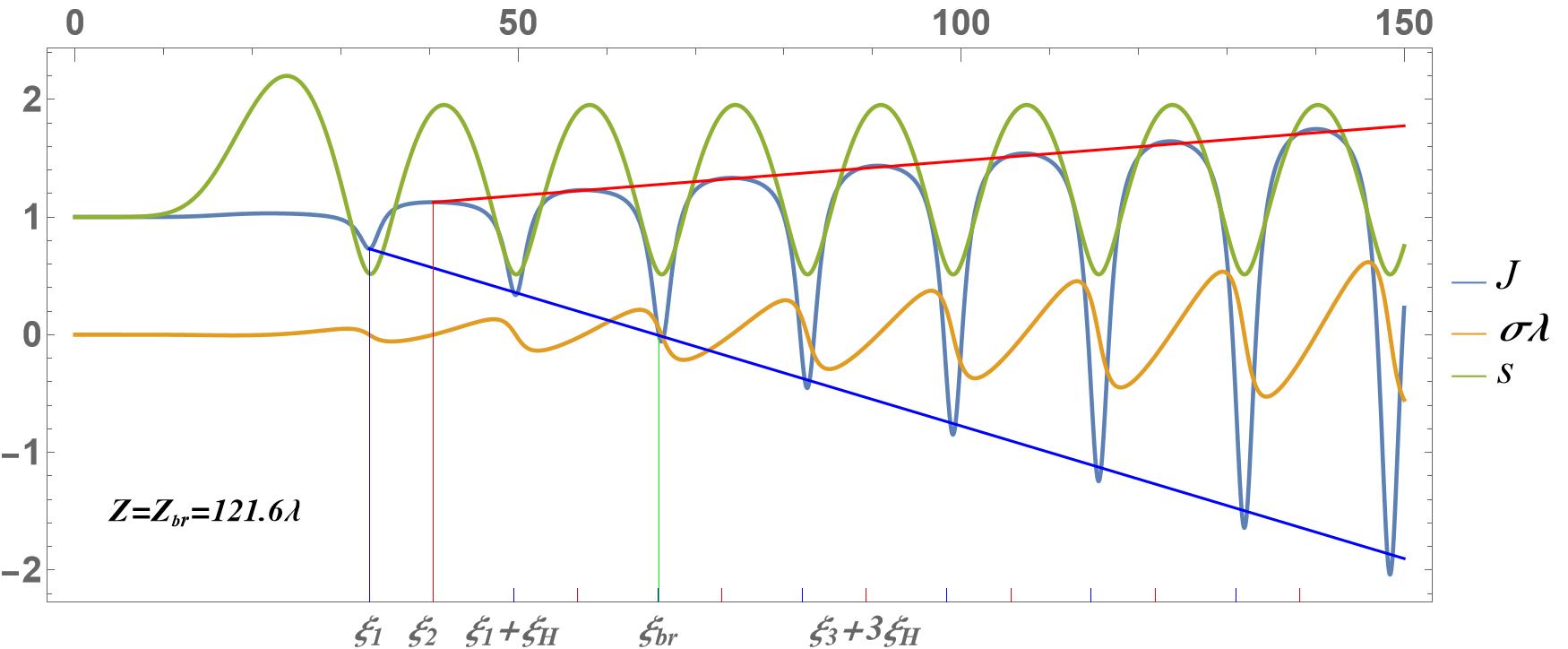} 
\caption{$\hat J,\hat\sigma$  vs. $\xi$ for $Z=Z_{br}\simeq121.6\lambda$ and input data as in Fig.  \ref{Worldlinescrossings-new}.
}
\label{graphs2'}
\end{figure}

The  map  $\hat \bx_e(\xi,\cdot)\!:\!\bX\!\mapsto\!  \bx$ from Lagrangian to Eulerian coordinates is invertible,  and the HD  justified,  as long as 
\be
\hat J \equiv \left|\frac{\partial\hat \bx_e}{\partial \bX}\right|  \stackrel{(\ref{hatsol'})}{=} \frac{\partial\hat z_e}{\partial Z} > 0 
\ee
for all $Z$. The electron density diverges where $\hat J=0$, because
\be
n_e(t,z)\!=\!\left[ \frac{\hat\gamma\,\widetilde{n_{0}}}{\hat s \,\hat J}\right]_{(\xi,Z)=\big(ct\!-z,\hat Z_e(ct\!-z,z)\big)}.
\label{expln_e}
\ee
The identity  $\hat z_e\!\left[\xi\!+\!i\xiH(\!Z\!),\!Z\right]\!=\!\hat z_e(\xi,\!Z)$  holds for 
$i\!\in\!\NN$, $\xi\!>\!l$; differentiating  w.r.t. $Z$ and setting $\Phi \!\equiv\!\frac{\partial\xiH}{\partial Z}$ one  finds \cite{FioDeNAkhFedJov23}
\be
\hat J\!\left(\xi\!+\!i\xiH,Z\right)=\hat J(\xi,Z)-i\,\Phi (Z)\,\Delta'\!\left(\xi,Z\right)\!,
  \label{pseudoper}
\ee
so that (\ref{lin-pseudoper}) holds with   $b\equiv -\hat\Delta' \frac{\partial \log\xiH}{\partial Z}$, 
$a\equiv \hat J\!-\!\xi b$, see  Fig. \ref{graphs2'}. 
Since $\Delta'$ changes sign twice in each period, WB at large $\xi$ (and $t$) is unavoidable  \cite{Daw59}, unless $\widetilde{n_0}\!=$const. Via (\ref{pseudoper}) we can extend our knowledge of  $ \hat J$ from $[l, l\!+\!\xiH[$ 
to all $\xi\ge l$ and determine the number  of periods leading to the first WB. 
Outside the {\it future Cauchy developments} of WBs the HD keeps valid;
inside them the dynamics must be modified, see \cite{Fionew}.

Differentiating   (\ref{heq1}-\ref{heq2}) w.r.t. $Z$ we find that $\hat J$, $\sigma\!\equiv\!\frac{\partial \hat s}{\partial Z}$ fulfill
\be
\ba{ll}
\hat J'=-\frac {1\!+\!v}{\hat s^3}\hat\sigma, \qquad  & \hat\sigma'=K\left(\check n\, \hat J\!-\! \widetilde{n_0}\right), \\[4pt]
\hat J(0,Z)=1, \qquad &\hat\sigma(0,Z)=0,\ea      \qquad          \label{basic} 
\ee
where $\check n(\xi,Z)\equiv\widetilde{n_0}\left[\hze(\xi,Z)\right]$.
Studying (\ref{basic}) one can find \cite{FioDeAFedGueJov22,FioDeNAkhFedJov23} sufficient conditions on $\widetilde{n_0},\Bep$ (formulated in terms of $\Delta_u,\Delta_d,n_u,n_d,...$, cf. Fig. \ref{fig1}a) to avoid
wave-breaking
during the laser-plasma interaction (WBDLPI), i.e. whereby $\hat J(\xi,Z)>0$ for all $Z$ and $\xi\in[0,l]$, so that the first WB occurs  for $\xi\!>\!l$ and  can be controlled via (\ref{pseudoper}).

\section{WFA of (self-)injected electrons}
\label{WFA}

If a test $e^-$ is injected  in the PW at 
$\big(\hat z_i,\hat s_i\big)_{\xi=\xi_0}\!=\!(z_{i0}, s_{i0})$,  with $\xi_0\!>\!l$, 
$s_{i0}\!>\!0$, $\hbup_i(\xi_0)=0$
then \cite{FioDeNAkhFedJov23}   its $\hat z_i,\hat s_i$ evolve after
\begin{IEEEeqnarray}{l}
\hat z_i'= \displaystyle\frac {1\!-\!\hat s_i^2}{2\hat s_i^2}, \quad 
\hat s_i'(\xi)=K\!\left\{\!
\widetilde{N}\!\left[\hat z_i(\xi) \right] \!-\! \widetilde{N}\!\left[\hat Z_e\!\left(\xi,\!\hat z_i(\xi) \right) \right] \!\right\}\!.\qquad                 \label{heq-test}
\end{IEEEeqnarray}
Along the density plateau  (\ref{heq-test}b) reduces to $\hat s_i'=M\Delta$. Hence
\be
\hat s_i(\xi)
=\delta s+s(\xi),\quad\hat z_i(\xi)=z_{i0}+\!\!\int^\xi_{\xi_0}\!\!  \frac{dy}2\left[\frac {1}{\hat s_i^2(y)}\!-\!1\right],  \label{test-motion}
\ee
if $z_{i0}\!\ge\! z_q\!\equiv\! z_s\!+\!\DM(z_s)$.
 Here  $s(\xi)$ is the $s(\xi;n_0)$ of section \ref{specialcase}, and \ $\delta s\equiv s_{i0}\!-\!s(\xi_0)$. \
If the \ {\bf trapping condition} \cite{FioDeNAkhFedJov23} \ $s_i^m\equiv \smm\!+\!\delta s < 0$ \ is fulfilled, then there is $\xi_f\!>\!\xi_0$ such that $\hat s_i(\xi_f)=0$, $s'(\xi_f)<0$ ($\xi_f$ corresponds to $t_f=\infty$);  the $e^-$ is 
  {\bf trapped in a trough of the PW and accelerated}:
for $\xi\simeq \xi_f$ we have $\hat s_i(\xi)\simeq \! \left|s' (\xi_f)\right|(\xi_f\!-\!\xi)=\! M\left|\Delta (\xi_f)\right|(\xi_f\!-\!\xi)$, 
\be
\hat z_i(\xi)\, \simeq   \,
\frac {1}{2\left[ M\Delta(\xi_f)\right]^2 (\xi_f\!-\!\xi)}
\: \stackrel{\xi\to \xi_f}{-\!\!\!-\!\!\!\longrightarrow} \: \infty. 
\label{approx}
\ee
Solving (\ref{approx}) for $\xi_f\!-\!\xi$ we can express  $\hat s_i,\hat \gamma_i$ in terms of $ z_i$:
\be
\gamma_i=\frac{1}{2 s_i}+\frac{s_i}2
\:\simeq \: F \: \frac{z_i}{\lambda}\: \stackrel{ z_i\to \infty}{-\!\!\!-\!\!\!-\!\!\!\longrightarrow} \: \infty, \label{s_i^m<0|s_iz_i}
\ee
$F\!\equiv\! M\lambda \left|\Delta(\xi_f)\right|$.
Hence, in this simplified model trapped test $e^-$ {\it cannot dephase}
(as the phase velocity of the PW is $c$) and {\bf their  energy grows proportionally to the travelled distance}. 
Of course, (\ref{s_i^m<0|s_iz_i}) is reliable in the interval $0\le z_i\le \znd$ where pulse depletion can be neglected.
Fixed  $z_i,n_0$, if  $\xi_0,z_0,s_0$ lead to  $\delta s=-1$, then   $|\Delta(\xi_f)|=|\Dm|$, and $\gamma_i$ is maximized:
\be
\gamma_i(z_i,n_0) \: \simeq \: \sqrt{ j(\nu)}\:\:  
z_i/\lambda;         \label{gamma^M(z_i)}
\ee
here $j(\nu)\equiv 8\pi^2\nu\big[\bar h(\nu)\!-\!1\big]$ is proportional to the maximal longitudinal electric field due to the PW along the density plateau, and $\bar h(\nu)$ is the final electron energy transfered by the pulse if $\widetilde{n_0}(z)\!=\!n_0$, vs. 
$\nu\!\equiv\!n_0/n_{cr}$. 
Now we can formulate our 4-steps optimization procedure.

\subsection*{Step A: Computing $\bxiH(\nu)$, $\bar h(\nu)$, $j(\nu)$ for the given pulse}
\begin{figure}
\includegraphics[width=9cm]{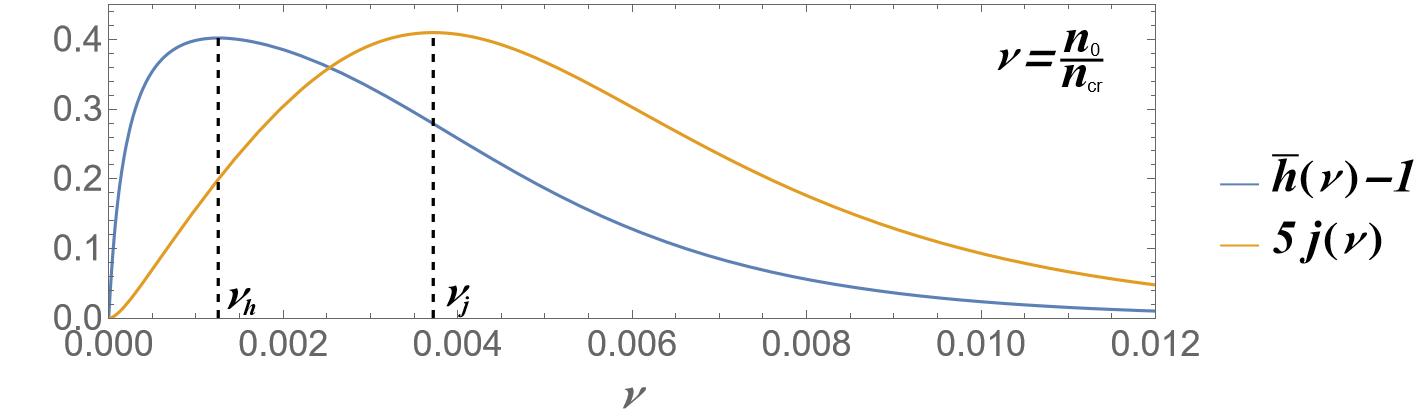} 
\caption{The energy gain per electron $\bar h\!-\!1$ and $j$ vs. the  
density $n_0$.}
\label{graph_bh-j_vs_nu}
\end{figure}

This is done  in few seconds (interpolating listplots of few hundreds points) using e.g. {\it Mathematica}. 
In Fig. \ref{graph_bh-j_vs_nu} we plot  $\bar h(\nu),j(\nu)$ and their 
maxima  $\nu_h,\nu_j$ for the pulse of Fig. \ref{graphsb}.a.

\subsection*{Step B: Optimal choice for the plateau density $n_0$}
If the plasma longitudinal  thickness $z_i$ available for WFA fulfills $z_i\le \znd(\nu_j)$, the best choice to maximize
$\gamma_i$ is $\nu=\nu_j$:
\be
\gammaM_i(z_i)\simeq\sqrt{j(\nu_j)}\, z_i/\lambda.   \label{gamma^MM(z_i)}
\ee

\subsection*{Step C: Optimal  $\widetilde{n_0}$ linear down-ramp for self-injection, LWFA}

By low plasma density,  significant 2-particle 
scatterings are rare, hence we  keep treating the plasma as collisionless. 
For all $Z$ the $Z$ $e^-$ keep making a comoving transverse layer feeling the same average $E^z$ due to the $x,y$-independent charge distribution.
Now assume $\widetilde{n_0}(z)$  decreases in some interval $\I_d\equiv[z_b,z_s]$, and
let $(\xi_{br},Z_{br})$ be the pair $(\xi,\!Z)$ with $Z\!\in\!\I_d$ and the smallest $\xi$ 
such that  $\hat J(\xi,\!Z)\!=\!0$.
 For $\xi\!>\!\xi_{br}$ a bunch of $Z\!\sim\! Z_{br}$ electron layers start crossing each other, breaking the PW locally. $\hat P(\xi,\!Z_{br})$ fulfills (\ref{heq-test}),
which is no longer decoupled from the eq.s   of other $e^-$ layers.
The $Z_{br}$  layer  earliest crosses other ones; at each $\xi\!>\!\xi_{br}$ it overshoots a new  layer that up to $\xi$ has evolved via (\ref{heq1}) and contributed to the PW. It does for ever, as the mutual repulsive forces  push it forward and the crossed one backward; hence the $Z_{br}$ are the fastest electrons
injected and trapped  by the first WB in a trough of the PW.
Fixed $z_{i0}\!\ge\!z_q$, let $\xi_0\!>\!\xi_{br}$ be the `instant'  when $\hat z_e(\xi_0,\!Z_{br})\!=\! z_{i0}$.
For $\xi\ge\xi_0$
$(\hat z_i,\hat s_i)\!\equiv\!\big(\hat z_e(\cdot,\!Z_{br}),\hat s(\cdot,\!Z_{br})\big)$ is given by
(\ref{test-motion}) and has $s_i^m<0$. 
For simplicity we stick to linear downramps
\be
\widetilde{n_0}(z)=n_0\!+\!\Upsilon(z\!-\!z_s),\qquad z_b\le z\le z_s,
\label{LinDownramp}
\ee
$\Upsilon\!<\!0$; abbreviating  $n_{b}\!\equiv\!\widetilde{n_0}(z_b)$, $n_{br}\!\equiv\!\widetilde{n_0}(Z_{br})$,
 $\delta Z\!\equiv\!z_s\!-\!Z_{br}$, and using the relations  $z_b\!=\!Z_{br}\!+\!\Dm\!(n_{br})$,
$\Upsilon\!=\!\frac{n_0\!-\!n_b}{z_s\!-\!z_b}\!=\!\frac{n_0\!-\!n_{br}}{\delta Z}$,
we can adopt $\Upsilon,z_b$ or $n_{br},\delta Z$ as parameters. We bound their range requiring that: there is no WBDLPI;
$P(\xi_0)$ is in the upper part of the cycle of Fig. \ref{graphsb}.c, i.e.
 at $\xi\!=\!\xi_0$ the $Z_{br}$ electrons cross plateau ones having negative displacement $\Delta$ and velocity $\Delta'$.
Abbreviating  $r(n_0)\!\equiv\!\frac{\bar h(n_0)}{\sqrt{\bar h^2(n_0)\!-\!1}}\!-\!1$, 
we can approximate \cite{Fionew} the latter requirement by the inequalities
\be
\frac12 \bxiH(n_0) \ge \delta Z\,r(n_0) \ge \frac14 \bxiH(n_{br})-\DM(n_{br})
\label{deltaZbounds}
\ee
and \ \ $\delta s\equiv s_{i0}\!-\!s(\xi_0)$  \ \ as a function of  $n_{br},\delta Z$ by
\begin{IEEEeqnarray}{c}
s_{i0}=C-\sqrt{C^2\!-\!1}, \quad \nonumber
C\!=\!Kn_{br} (z_{i0}\!-\!Z_{br})\zeta\!+\!\bar h(n_{br}), \qquad
\end{IEEEeqnarray}
\begin{IEEEeqnarray}{c}
\Xi=\frac12 \bxiH(n_0)\!-\! \delta Z\,r(n_0),
\quad\sM\!=\!\bar h(n_0)\!+\!\sqrt{\bar h^2(n_0)\!-\!1},\qquad\\
 \zeta=\frac{\Xi}2\!\left(\!1\!-\! \frac {M\Xi^2}{3\sM^3}\!\right)\!\left(\!1\!-\!\frac {1}{\sM^2}\!\right),
\quad  s(\xi_0)=\sM\!-\!\frac{M\Xi^2}4\!
\left(\!1\!-\!\frac {1}{\sM^2}\!\right)\nonumber
\end{IEEEeqnarray}
The third step consists in 
making $\delta s$ as close as possible to -1, so that  (\ref{gamma^M(z_i)}) applies,
by varying $n_{br},\delta Z$.

\subsection*{Step D: choosing the $\widetilde{n_0}$  up-ramp so as to prevent earlier WB}

We pick one of the $\infty$-ly many \cite{FioDeAFedGueJov22,FioDeNAkhFedJov23} 
$\widetilde{n_0}(z)$ growing from 0 to $n_b$ 
in an (as short as possible)  interval $0\!\le\!z\!\le\!z_b$  and preventing WB for $\xi\!<\!\xi_{br}$; that $\widetilde{n_0}(z)\simeq O(z^2)$ 
helps.

\section{Discussion and conclusions}
\label{Conclu}

\begin{figure}
\includegraphics[width=9.2cm]{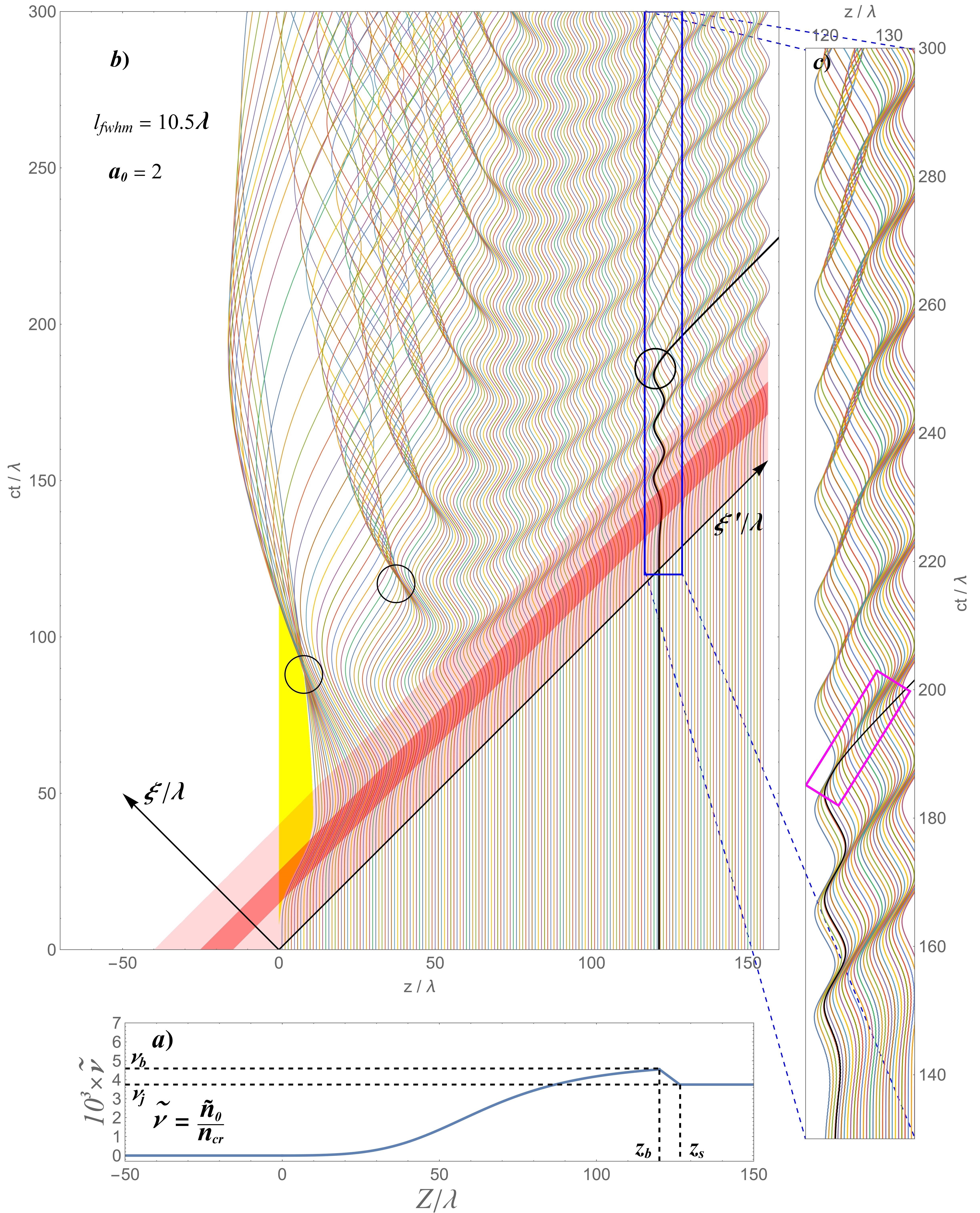}
\caption{a) \  Optimal initial plasma density $\widetilde{n_0}(Z)$ for the 
pulse of Fig. \ref{graphsb}.a: $n_0 \!=\!n_0^j\!=\!n_{cr}/267$,  \
$n_{b}\!=\!1.21\times n_0^j$, \ $z_b \!=\! 120\lambda$, \
$z_s\!-\!z_b \!=\! 6.6\lambda$. \
b) \  Projections onto the $z,ct$ plane of the corresponding WLs (in Minkowski space)
of the $Z$ electrons 
 for $Z=0,\lambda,...,156\lambda$. \ We have studied the down-ramp $Z$ electrons more in detail, determining their
 WLs for  $Z=120\lambda,120.1\lambda,...,140\lambda$: 
in \ c)   \ we zoom the blue box of a). 
Here: $\xi'\equiv ct\!+\!z$; in the dark yellow region only ions are present; we have painted pink, red the support of $\Bep(ct\!-\!z)$  (considering $\Bep(\xi)\!=\!0$ outside $0\!<\!\xi\!<\!40\lambda$) and the region where the modulating intensity 
is above half maximum, i.e. $-l'/2\!<\!\xi\!-\!20\lambda\!<\!l'/2$, with $l'
=10.5\lambda$; the pulse  can be considered ES [cf. (\ref{Lncond'})] if we consider  some $l_r\le 27\lambda$ as the pulse length, instead of $l=40\lambda$.
}
\label{Worldlinescrossings-new}       
\end{figure}

We will present derivation and details of the above 4-steps procedure  in \cite{Fionew}.
Here we apply it to tailor the plasma density   to  the laser pulse of Fig. \ref{graphsb}.a. 
We depict the resulting $\widetilde{n_0}(z)$ and
the corresponding $e^-$ WLs  in Fig. \ref{Worldlinescrossings-new}. The first WB
(rotated magenta box in Fig. \ref{Worldlinescrossings-new}c) occurs at $\xi_{br}\simeq
66\lambda$ and involves the $Z\!\sim\! Z_{br}=121.6\lambda$ down-ramp electrons; the $Z\!=\!Z_{br}$ ones  nearly fulfill (\ref{gamma^M(z_i)})
(with $\delta s\!=\!-1.01$) and  (\ref{gamma^MM(z_i)}), hence 
have the largest possible WFA factor  (\ref{s_i^m<0|s_iz_i}),  $F\!=\!\sqrt{ j(\nu_j)}\!=\!0.286$;  we have plot their WL black. 
This WB is causally disconnected from the earliest WBs (encircled in Fig. \ref{Worldlinescrossings-new}b) involving up-ramp $e^-$s.
More generally, fixed any plateau density, we can apply  steps A,C,D  to 
maximize WFA of the $e^-$s injected in the PW by the first WB; picking $n_0\!=\!n_{cr}/267$,
the WFA factor  $F$ is $1.36$ times that found in \cite{FioDeNAkhFedJov23} for optimal input data of \cite{BraEtAl08}. 

Now assume the pulse is not a plane wave, but cylindrically symmetric around  $\vec{z}$  with a {\it finite} spot radius $R$, i.e. at $t=0$ \ \
$\bE=\Be^{{\scriptscriptstyle\perp}}\!(-z)\,\chi(\rho),\:\: \bB=
\bk\!\times\!\bE$,
 where $\rho^2\!=\!x^2\!+\!y^2$ and $\chi(\rho)\ge 0$ is 1 for $\rho\le R$ and 
goes fast to 0 
as $\rho\to\!\infty$.  By causality,  our results hold strictly inside the causal cone (of axis $\vec{z}$, radius $R$) trailing the pulse and approximately in a neighbourhood thereof. In particular, if the pulse has maximum
at $\xi=\frac l 2$, and
\be
R\:>\: \xi_{br}-\frac l 2,\quad R\:\gg\:
\frac{a_0\lambda}{2\pi}\left[\bar h\!+\!\sqrt{\bar h^2\!-\!1}\right]\simeq
 |\Delta\bx^{{\scriptscriptstyle \perp}}_{e{\scriptscriptstyle M}}| 
\ee
(this excludes the {\it bubble regime} \cite{RosBreKat91,MorAnt96,PukMey2002
}), then
 the $\bX\simeq(0,\!0,\!Z_{br})$ $e^-$ are injected in the PW as above and trail that cone with the same maximal WFA, as far as the pulse is not depleted. 
Checking our results via PIC simulations is welcome.


\end{document}